\documentstyle[preprint,aps,prb,tighten,epsf]{revtex}

\begin{document}
\draft

\title{Transport properties of quasiparticles with fractional
exclusion statistics}

\author{I. V. Krive$^{1,2}$ and E. R. Mucciolo$^{1}$}

\address{$^{1}$Departamento de F\a'{\i}sica, Pontif\a'{\i}cia
Universidade Cat\'{o}lica do Rio de Janeiro, Caixa Postal 38071,
22452-970 Rio de Janeiro, Brazil}
\address{$^{2}$B.I. Verkin Institute for Low Temperature Physics and
Engineering, Kharkov, Ukraine}

\date{August 20, 1998}

\maketitle

\begin{abstract}
We consider the ballistic transport of quasiparticles with exclusion
statistics through a 1D wire within the Landauer-B\"{u}ttiker
approach. We demonstrate that quasiparticle transport coefficients
(electrical and heat conductance, as well as thermopower) are
determined by the same general formulae as for particles with normal
statistics. By applying the developed formalism to the ballistic
transport of fractional charge it is shown that for a wire in contact
to quasiparticles reservoirs the transport coefficients depend on the
fractional charge. Specific features of resonant tunneling of
quasiparticles are discussed.
\end{abstract}

\narrowtext

\section{Introduction}

Exclusion statistics was proposed \cite{Haldane91} as a
phenomenological description of excitations with mixed statistical
properties (intermediate between fermions and bosons) in many-body
systems. It is based on the assumption that the change of the number
of available one-particle states in the system, $\Delta d$, at a given
volume and at fixed boundary conditions depends {\em linearly} on the
change of the number of quasiparticles $\Delta N$: $\Delta d = -
g\Delta N$. Here $g$ is the statistical parameter which is assumed to
be a rational number, with $g=0$ and $g=1$ corresponding to the normal
bosonic and fermionic statistics, respectively. This generalized
Pauli's principle results in a $g$-dependent statistical weight for
the ideal gas of quasiparticles (usually referred as {\em
exclusons}). The standard methods of statistical quantum mechanics
allow one to derive the equilibrium distribution function of exclusons
\cite{Wu94}
\begin{equation}
f_g(\varepsilon) = \left[ y\left( \frac{\varepsilon-\mu_g}{T} \right)
+ g\right]^{-1},
\label{eq:distribution}
\end{equation}
where $T$ is the temperature (we consider units where $k_B = 1$),
$\mu_g$ is the chemical potential of the ideal gas, and $y(x)$ is the
real positive solution of the algebraic equation
\begin{equation}
y^{g}(1+y)^{1-g} = e^{x}.
\label{eq:implicit}
\end{equation}
Formally, exclusons can be regarded as composites of fermions and
bosons since their thermodynamic potential can be decomposed into a
sum of fermionic and bosonic potentials weighted by fractional
coefficients,\cite{Iguchi97} namely,
\begin{eqnarray}
\Omega_g & = & -T\sum_{j}\ln\left[\frac{1+(1-g)f_g(\varepsilon_j)}
{1-gf_g(\varepsilon_j)}\right] \nonumber \\ & = & g\Omega_f +
(1-g)\Omega_b,
\label{eq:thermopot}
\end{eqnarray}
where $\Omega_{g,f,b}$ denotes the grand thermodynamic potential for
exclusons, fermions, and bosons, correspondingly, and $\mu_g = g\mu_f
+ (1-g)\mu_b$.

In spite of its phenomenological footing, exclusion statistics proved
to be correct for quasiparticles in certain exactly solvable 1D models
\cite{Kohmoto??} and, more importantly, it is the statistics of
recently observed fractionally charged excitations in fractional
quantum Hall systems,\cite{Saminadayar97} with the statistical
parameter determined by the Landau level filling factor.

Already for two decades, fractional charge is a theoretically
well-grounded concept, both in quantum field theory and in solid state
physics (for reviews, see Ref. \onlinecite{Niemi86Krive87}). It is
known \cite{Hatsugai92} that the requirement of gauge invariance
forces the gas of fractionally charged particles at equilibrium to
obey fractional statistics. From a phenomenological point of view,
exclusion statistics is a reasonable candidate for the required
anomalous statistics of fractionally charged quasiparticles. Thus, the
study of thermodynamic and transport properties of {\em fractionally}
charged exclusons can be more than simply an academic exercise.

In Ref. \onlinecite{Krive96} the persistent current of a 1D ring of
fractionally charged exclusons was calculated. It was shown that at
low temperatures the fractional charge, masked by fractional
statistics, does not show up in the Aharonov-Bohm oscillations.
However, the ``high-temperature'' properties of the system do depend
on fractional charge. Unfortunately, this region is hardly accessible
to present experiments since the oscillating current itself is
exponentially small. The purpose of the present paper is to go beyond
thermodynamics and consider transport properties of exclusons. We show
within the Landauer-B\"{u}ttiker formalism that the transport
coefficients for exclusons are determined by the same general formulae
as for the particles with normal statistics, provided the leads
representing particle reservoirs are also described by exclusion
statistics. In this case, even for an impurity-free 1D wire the
transport coefficients depend on the statistical parameter. It is
argued however that for an ideal wire the above dependence disappears
when one considers ``normal'' particle reservoirs.

\section{Conductance}

Let us consider an effectively infinite 1D wire of exclusons with a
scattering potential described by the transmission probability
$T_t(\varepsilon)$. We will assume at first that the wire is connected
to reservoirs of {\em exclusons} in equilibrium at temperature $T$ and
chemical potential $\mu_g$. In this case the generalization of
Landauer's formula for conductance is straightforward: For exclusons
with fractional charge $q_m=e/m$ ($m$ is an integer) and statistical
parameter $g=1/m$ it reads
\begin{equation}
G_m = \frac{G_0}{m^2} \int_{0}^{\infty} d\varepsilon\ T_t(\varepsilon)
\left[ -\frac{\partial f_g(\varepsilon)}{\partial\varepsilon} \right],
\label{eq:conductance}
\end{equation}
where $G_0=(2s+1)e^2/h$ is the conductance quantum ($s$ is the spin of
exclusons) and the distribution function is defined by
Eqs.~(\ref{eq:distribution}) and (\ref{eq:implicit}). To proceed we
have to specify the properties of the scattering potential in the
vicinity of the Fermi energy. If transport of exclusons through the
barrier is nonresonant, the transmission probability at low
temperatures ($T\ll\mu_g$) can be approximated by an energy
independent factor $T_t(\mu_g)$. Then, Eq.~(\ref{eq:conductance}) is
reduced to a simple formula
\begin{equation}
G_m = \frac{G_0}{m}T_t(\mu_g).
\label{eq:condsimple}
\end{equation}
In particular, for a perfect wire ($T_t=1$) the conductance of
fractionally charged exclusons $G_g = gG_0$ coincides with that for a
homogeneous Luttinger liquid (LL) of integer charge particles with
correlation parameter $g=1/m$. However, in distinction from the LL
case, exclusons are supposed to be noninteracting particles and
therefore their scattering by a local potential can be described by a
non-renormalized transmission probability $T_t(\mu_g)$. Thus the
conductance is finite and temperature independent up to temperatures
of the order of $\mu_g$. Here a clarifying comment is needed. We are
considering a gas of exclusons in a grand canonical ensemble, so
$\mu_g$ is an input parameter, which is assumed to be positive. It
means that for any $g\neq 0$ there exists a low temperature region
where $T\ll\mu_g$. Yet, for $g\ll 1$ the chemical potential $\mu_g =
g\mu_f - (1-g)\mid\mu_b\mid>0$ can also be small and the only region
realizable is the ``high temperature'' one $T\gg\mu_g$. In this case
the conductance will be sensitive to the low energy dependence of the
transmission coefficient. For a perfect wire ($T_t=1$) the
corresponding conductance will read $G_{m\gg 1}(T\gg\mu_g) \simeq
G_0/m\ln m$. In what follows we will assume that even for large (but
finite) $m$ the excluson chemical potential is still the largest
energy scale in the problem.

The case of resonant tunneling of exclusons reveals far more
interesting transport properties than the homogeneous case. To show
why, we will approximate the resonant transmission coefficient
$T_t(\varepsilon)$ at energies in the vicinity of the Fermi level
($\varepsilon\simeq\mu_g$) by the Breit-Wigner form
\begin{equation}
T_t^{(r)}(\varepsilon) = \frac{\Gamma^2}{(\varepsilon-\Delta)^2 +
\Gamma^2},
\label{eq:breitwigner}
\end{equation}
where $\Delta$ and $\Gamma$ are the position and the width of the
resonance level, respectively. The resonance tunneling implies that
$\mu_g^{(r)} = \Delta$ and the corresponding conductance at $T=0$
coincides (at it should be) with that for a perfect channel. At finite
temperature the resonant conductance takes the form
\begin{equation}
\label{eq:7}
G_m^{(r)}(T) = \frac{G_0}{m^2} \int_{-\infty}^{\infty} dx
\frac{(\Gamma/ T)^2}{x^2+(\Gamma/T)^2} \left[ -\frac{\partial
f_g(x)}{\partial x} \right].
\label{eq:resonantcond}
\end{equation}
The above expression determines the temperature dependence of the
resonance peak height at $T\ll\mu_g$.  Let us place the question - How
does the peak height depend on the statistical parameter $g=1/m$? If
$g$ is not small the resonance conductance is qualitatively the same
as that for fermions ($m=1$),
\begin{equation}
G_m^{(r)}(T) \simeq \frac{G_0}{m} \left\{ \begin{array}{lr} 1
-(A_m/m) (T/\Gamma)^2 , & T \ll\Gamma \\ B_m (\Gamma/T), & T
\gg\Gamma \end{array} \right.,
\label{eq:asymptotics}
\end{equation}
where
\begin{equation}
A_{1/g} = \int_{-\infty}^{\infty} dx\ x^2 \left[ -\frac{\partial
f_g(x)}{\partial x} \right]
\label{eq:A_m}
\end{equation}
and
\begin{equation}
B_{1/g} = \pi g \left[ -\frac{\partial f_g(x)}{\partial x}
\right]_{x=0}.
\label{eq:B_m}
\end{equation}
These coefficients can be evaluated as follows. With the help of
Eq.~(\ref{eq:implicit}) the derivative of the distribution function
can be expressed in the form
\begin{equation}
f_g^{\prime}(x) = - \frac{y(x)[1+y(x)]}{[g+y(x)]^3}
\label{eq:derivative}
\end{equation}
Inserting this expression into Eq.~(\ref{eq:A_m}) and changing the
integration variable from $x$ to $y$ one finds for any $g\neq 0$ that
the coefficient $A_{1/g}$ takes the same value as for fermions,
\begin{equation}
A_{1/g} = \int_0^{\infty}dy\frac{[g\ln y + (1-g)\ln(1+y)]^2} {(g +
y)^2} = \frac{\pi^2}{3}.
\label{eq:A_mindep}
\end{equation}
This is a remarkable fact - the coefficient $A_{1/g}$ does not depend
on the statistical parameter. In the next section it will be shown
that this very coefficient determines the heat conductance of
exclusons. The numerical coefficient $B_{1/g}$ on the other hand, does
depend on $g$ and takes the values: $B_1= \pi/4$, $B_2 =
4\pi/5\sqrt{5}$, $\ldots,$ $B_{m\gg 1}\sim m/\ln^{2}m$.

For decreasing values of the statistical parameter the temperature
dependence of the resonance conductance acquires new qualitative features.
Namely, the crossover from low-T to high-T behavior is transformed
into a wide region $\Gamma\ll T\ll m\Gamma$ where $G_m(T)$ 
for $m\gg 1$ depends
almost {\em linearly} on temperature (see Fig.1). To study this region
analytically it is tempting to use the bosonic limit. By expanding the
distribution function of exclusons in the vicinity of the bosonic
statistics ($g=0$) it is easy to show that first two terms of
$1/m$-expansion of resonant conductance, Eq.~(\ref{eq:resonantcond}),
take the form
\begin{equation}
G_{m\gg 1}^{(r)}(T) \simeq \frac{G_0}{m}\left\{1
-\frac{1}{m}\left[\gamma \psi^{\prime}(\gamma)-
\frac{1}{2\gamma}-1\right]\right\},
\label{eq:intermed}
\end{equation}
where $\gamma \equiv \Gamma/2\pi T$ and $\psi^{\prime} \equiv d^2\ln
\Gamma(x)/dx^2$ (here $\Gamma(x)$ is the gamma
function\cite{Gradshteyn80}). For finite $m$ the above expression
holds formally until the term in the square brackets is much smaller
than $m$. This restricts the temperature region where
Eq.~(\ref{eq:intermed}) could be at least qualitatively correct to
temperatures $T\ll\Gamma_m\equiv m\Gamma/\pi$. According to
Eq.~(\ref{eq:intermed}), at low temperatures ($T<\Gamma$) the
resonance conductance is still determined by the low-$T$ asymptotics
of Eq.~(\ref{eq:asymptotics}). However, for $m\gg 1$ there is a wide
temperature region $\Gamma<T \ll\Gamma_m$ where conductance depends
{\em linearly} on temperature, $G_{m\gg 1}^{(r)}(T) \simeq (G_0/m) (1
- T/\Gamma_m)$. It is only at temperatures $T>\Gamma_m\gg\Gamma$ that
the resonance conductance switches to the ordinary high-$T$
asymptotics. Notice that Eq.~(\ref{eq:intermed}) is not the true
asymptotic expansion because all higher terms in $1/m$ are represented
by divergent integrals. The formula is not quantitatively reliable and
can be regarded at most as an estimate. We presented it only to show
how the intermediate linear dependence on temperature appears in the
crossover region of Eq.~(\ref{eq:asymptotics}). It is worth noticing
that the widening of the crossover region is a specific feature of the
exclusion statistics for resonant tunneling. The above discussed
linear-$T$ regime is pronounced only for very small values of
statistical parameter $g$.

The considerations above dealt with the transport of fractionally
charged quasiparticles through a wire connected to reservoirs
containing the particles of the same charge and statistics. Although
theoretically conceivable, such situation is difficult to realize
experimentally. A more natural experimental setup would be a 1D system
of exclusons connected to leads containing noninteracting
electron. Does the conductance of a perfect "excluson wire" connected
to electron reservoirs depend on statistical parameter $g$? To solve
this problem honestly one needs to introduce a microscopical model of
fractional charge transport. In the phenomenological approach
developed above it is reasonable to consider the case when the
"excluson wire" is a metallic state characterized by a Fermi energy
$\varepsilon_F^{(ex)}$. When an electron passes through the
(presumably perfect) boundary to the "excluson wire" it is converted
into $m$ particles with fractional charge $q_m =e/m$. Therefore, the
fractional charge in the expression for the quasiparticle current will
be canceled by the extra factor $m$ in the density of exclusons. At
low temperatures and provided that there is no backscattering on the
boundaries, a voltage applied to the electron reservoirs will result
in a current through the system which will not depend on the
fractional charge. Formally, the above statement can be proved by
mapping the excluson model onto a LL model \cite{Wu95} (the mapping is
exact at $T=0$ and for the impurity-free case) and by recalling that
the conductance of a LL constriction connecting to reservoirs of
noninteracting electrons is $G_0=(2s+1)e^2/h$.
\cite{Maslov95Ponomarenko95Shafi95}

\section{Thermoelectric Effects}

In the previous section we showed that the Landauer's formula for
conductance can be applied to particles with exclusion
statistics. Could we expand this claim to all transport coefficients?
Since this is not evident from a general point-of-view, we at first
derive the expressions for heat conductance $K_g(T)$ and
thermoelectric cross coefficients $\Lambda_g(T)$ using the approach
developed in Ref. \onlinecite{Sivam86}.

Knowing the explicit expression for the grand thermodynamic potential
of exclusons, Eq.~(\ref{eq:thermopot}), it is easy to show that the
entropy of the ideal gas of exclusons takes the form
\begin{eqnarray}
S_g & = & -\sum_{j} \left\{ f_g(\varepsilon_j) \ln f_g(\varepsilon_j)
     + [1-gf_g(\varepsilon_j)] \ln[1-gf_g(\varepsilon_j)]
     \right. \nonumber \\ & & \left. - [1+(1-g)f_g(\varepsilon_j)]
     \ln[1+(1-g)f_g(\varepsilon_j)] \right\},
\label{eq:entropy}
\end{eqnarray}
where $f_g(\varepsilon_j)$ is the distribution function of exclusons
obtained from Eqs~.(\ref{eq:distribution}) and (\ref{eq:implicit}). In
the limiting cases of $g=1$ and $g=0$, Eq.~(\ref{eq:entropy})
transforms into a well-known formula for the entropy of the ideal
fermion and boson gases, respectively. With the entropy of exclusons
at hand one can (following Ref. \onlinecite{Sivam86}) perform all the
transformations leading to the expressions for the heat conductance
and the coefficient $\Lambda_g(T)$. Using Eqs.~(\ref{eq:distribution})
and (\ref{eq:implicit}) one can show that the desired quantities are
determined by the same general formulae as for ordinary
particles,\cite{Sivam86}
\begin{equation}
K_g(T) = \frac{(2s+1)}{h} \int_0^{\infty} d\varepsilon
\frac{(\varepsilon -\mu_g)^2}{T} T_t(\varepsilon) \left[
-\frac{\partial f_g(\varepsilon)} {\partial\varepsilon} \right].
\label{eq:thermocond}
\end{equation}
A similar expression [but linear on $(\varepsilon-\mu_g)$] stands for
the thermoelectric coefficient $\Lambda_g$.

The analysis of Eq.~(\ref{eq:thermocond}) is straightforward. At low
temperatures, $T\ll\mu_g$, one can replace the transmission
probability by the energy-independent factor $T_t(\mu_g)$. Then, the
heat conductance reads
\begin{equation}
K_g(T) = \frac{1}{h} T A_{1/g} T_t(\mu_g),
\label{eq:thermosimple}
\end{equation}
where the numerical factor $A_{1/g}$ is defined by Eqs.~(\ref{eq:A_m})
and (\ref{eq:A_mindep}). Since the above coefficient does not depend
on the statistical parameter, the low-$T$ heat transport of exclusons
is determined by the same formulae as for free fermions. Analogously,
the thermoelectric coefficient of exclusons in the limit
$\mu_g/T\rightarrow\infty$ vanishes irrespective of the value of $g$,
\begin{equation}
\Lambda_g(T) \propto \int_0^{\infty}dy\frac{g\ln y + (1-g)\ln(1+y)}
{(g+y)^2} \equiv 0.
\label{eq:thermoelect}
\end{equation}
We remark that for bosonic statistics the limit $g\rightarrow 0$ in
Eq.~(\ref{eq:thermocond}) should be taken before $T\rightarrow 0$. For
the case of neutral bosons ($g=0,\mu_b = 0$) the resulting heat
conductance is sensitive to the energy dependence of the transmission
probability at low energies. For a perfect channel ($T_t=1$) and
spinless particles, we obtain
\begin{equation}
K_{g\rightarrow 0}(T,\mu_g\rightarrow 0) = \frac{\pi^2}{3}
\frac{1}{h}T,
\label{eq:thermoperfect}
\end{equation}
which again coincides exactly with that for fermions.

An important characteristic of metallic systems is the ratio between
the heat and electric conductances, $L(T)=K(T)/TG(T)$, known as the
Lorentz number. [For Fermi liquids, $L_0=(\pi^2/3)(1/e)^2$].
According to Eqs.~(\ref{eq:condsimple}) and (\ref{eq:thermosimple})
this quantity for exclusons, though temperature independent for
$T\ll\mu_g$, depends on the statistical parameter $L_g = L_0/g$. This
result coincides exactly with that for an infinite Luttinger
liquid.\cite{Kane96}

\acknowledgments

This work was supported by the Brazilian Agency CAPES. I.K. thanks the
Physics Department at PUC-Rio (Rio de Janeiro) for the hospitality.


\begin{figure}
\setlength{\unitlength}{1mm}
\begin{picture}(120,120)(0,0)
\put(10,10){\epsfxsize=12cm\epsfbox{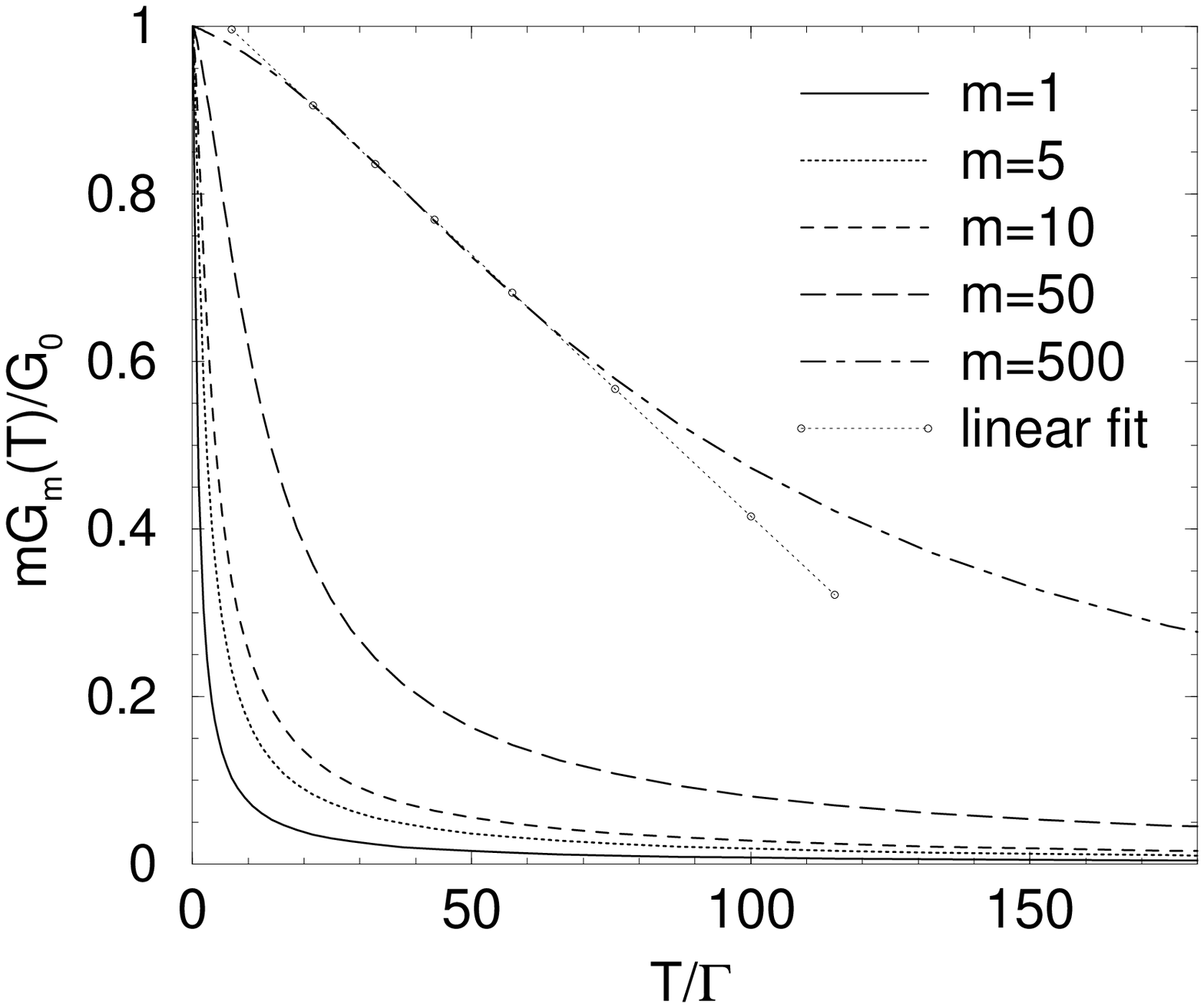}}
\end{picture}
\caption{Resonant conductance as a function of temperature for
different values of the statistical parameter $g=1/m$
[Eq.~(\ref{eq:7})]. The straight dot-dashed line is a linear fit to
the $m=500$ curve for $T<\Gamma_m$}
\label{fig1}
\end{figure}

\end{document}